\documentclass[10pt,conference]{IEEEtran}
\IEEEoverridecommandlockouts
\usepackage{cite}
\usepackage{amsmath,amssymb,amsfonts}
\usepackage{algorithmic}
\usepackage{graphicx}
\usepackage{textcomp}
\usepackage{xcolor}
\usepackage{subcaption}
\usepackage{url}
\def\BibTeX{{\rm B\kern-.05em{\sc i\kern-.025em b}\kern-.08em
    T\kern-.1667em\lower.7ex\hbox{E}\kern-.125emX}}
\begin{document}

\title{LEO Downlink Channel Model Revisited: Scattering Geometry-Inspired Derivation}

\author{\IEEEauthorblockN{ Kuan-Po Chiu}
\IEEEauthorblockA{\textit{Department of Electrical \& Computer Engineering} \\
\textit{University of Washington}\\
Seattle, WA, USA \\
kchiu5@uw.edu}
\and
\IEEEauthorblockN{Sumit Roy}
\IEEEauthorblockA{\textit{Department of Electrical \& Computer Engineering} \\
\textit{University of Washington}\\
Seattle, WA, USA \\
sroy@uw.edu}
}
\maketitle
\vspace*{-0.4in} 
\begin{abstract}
This paper presents a new derivation of LEO-to-ground receiver channel model to address a clear gap in the prior art: the lack of an appropriate geometry aware characterization of non LOS (NLOS) link model represented by the power spectral density (PSD). Specifically, the main contribution is a coherent derivation of the PSD from 1st principles that is able to reproduce results in prior art and explain the causal relationship of main PSD features to the propagation geometry parameters.
\end{abstract}


\section{Introduction}
The  downlink satellite-to-ground (S2G) channel  can be classified into either of the two categories (pure NLOS vs combination of LOS \& NLOS)
\begin{align}
    &h_{leo, dl}(t, \tau) \nonumber\\&= 
    \begin{cases}
        \sqrt{\frac{K}{K+1}} h_{\text{LOS}}(t) + \sqrt{\frac{1}{K+1}} h_{\text{NLOS}}(t, \tau) & \text{LOS-NLOS} \\[8pt]
        h_{\text{NLOS}}(t, \tau) & \text{NLOS}
    \end{cases}
\end{align}
where  $K > 0$  represents the power ratio in the LOS component relative to NLOS. {\em Critically, both LoS and NLoS channel components are a function of the elevation angle}. The LoS component captures the instantaneous Doppler shift frequency due to satellite motion (manifests itself as a spike in spectrum domain) while for NLoS components, the spatio-temporal distribution of the (secondary) scatterers in azimuth \& elevation around the receiver determines the shape and boundary of resulting power spectral density (PSD). LEO-to-ground propagation is subject to rapidly changing Doppler shifts due to significant (time-varying) relative motion of LEO satellites in LOS conditions. Nia and Mark\cite{doppler} developed a detailed model to predict the instantaneous Doppler shift as a function of time (equivalently, satellite location on its trajectory). 

The primary motivation for this work is to {\em revisit model based derivation of LEO S2G downlink channel\footnote{Hass\cite{Aeronautical} has adequately addressed the uplink scenario from a ground node to an airborne platform that employs a directional antenna; the resulting stochastic channel power spectral density (PSD) have been thoroughly examined.} to fill existing gaps} in understanding of causal relation between the propagation/scattering geometry and resulting Doppler PSD for non-LOS (multipath) scenarios. 
When the terrestrial receiver is located close to built (i.e. urban/suburban location) or other natural environments (trees/foliage) that act as potential nearby obstructions, it leads to  significant secondary scattering as shown in Fig.~\ref{fig1} that determines the nature of resulting short time-scale fading. Further, newer generation of LEO satellites employ larger phased arrays onboard capable of downlink beamforming that further impacts the spatio-temporal multipath delay profile experienced by a terrestrial client. 
%Finally, in nearly all prior efforts, the ground node (UE) is assumed to be stationary/fixed in the global coordinates; with the advent of direct-to-cell satellite features (e.g Starlink's recent announcement of beta-test of such a feature to all T-Mobile customers), ground user mobility must also be factored into statistical characterization of downlink channel. 

\section{Literature Review \& Contributions}
%Typical LEO satellite orbits with inclination relative to earth's equatorial plane are elliptical.
We briefly review how NLoS channel models in prior art have accounted for multipath profile based on the elevation angle $\beta_{ele}$ as in Fig.~\ref{fig1}. Aulin \cite{aulin} 
Janaswamy~\cite{3dspheroid} and Alsehaili et al.~\cite{3dellipse} all derive the PSD from first principles, assuming a probability density function(PDF) of scatterers in 3D space.  However, these are all 
targeted to {\em near terrestrial} propagation from an elevated base station to ground users, and is not applicable to the LEO satellite-to-ground (S2G) scenario, where capturing the impact of scatterer geometry around the ground user (receiver) as a function of the LOS direction (or equivalently the elevation angle $\beta_{ele}$) in the resulting is the primary concern.  Aulin's derivation assumes a spherical scatter geometry, i.e. invariant with respect to $\beta_{ele}$, while 
Janaswamy~\cite{3dspheroid} assumes a 3-D volumetric geometry of scatterers that varies with altitude and  Alsehaili et al.~\cite{3dellipse} ellipsoidal volumetric scattering models. 

The prior art on non-terrestrial to ground models includes Liu et al.\cite{fingerprint} that presents a PSD approximation for various elevation angles but without any derivation from first principles that provides insight into desired $\beta_{ele}$  dependence. Zhao et al.~\cite{53g} derives the PSD for various azimuth and elevation PDFs using a numerical approach that provides some useful qualitative insight but a clear relationship between the PSD and the LoS elevation angle is still missing. 
Newhall and Reed~\cite{a2g} derive the joint PDF at a given elevation angle using an ellipsoid scatter model for air-to-ground channel. While relevant, this model is not directly applicable for a LEO-to-ground scenario, as the distances between the transmitter and receiver is significantly greater compared to the plane-to-ground configuration. 
 
%\cite{3dspheroid} targets terrestrial downlink propagation (elevated BS to UE), while~\cite{3dellipse} assumes the signal source (base station) at the ellipsoid’s focus, with contributions from near and far scatterers. 
In summary, there does not exist an adequate scatterer-geometry inspired derivation of the S2G downlink channel model.

\section{Proposed Doppler PSD Model of S2G Channel}
% \subsection{3D Delay Spread Considerations}
\subsection{Propagation Model}
This study investigates LEO S2G communication, with typical orbital altitudes of 500 to 1500 km. To investigate the multipath characteristics of the S2G channel, our proposed model is based on the following assumptions:
\begin{itemize}
    \item The ground receiver is equipped with an isotropic antenna. 
    \item All multipath components arrive at the receiver with equal power, i.e all scatterers have the same scattering coefficient but uniform random phases. 
    \item The incident waves arriving at each scatterer can be approximated as planar waves, implying that the angle of arrival (AoA) to each scatterer is the same across all multipath components. This assumption holds under the condition of the large satellite-to-ground distance.
    \item Only multipath signals experiencing a single bounce due to scatterers nearby receiver are considered.

\end{itemize}

\subsection{General 3-D Autocorrelation Function}
\begin{figure}[!bp]
    \centerline{\includegraphics[width=1\linewidth]{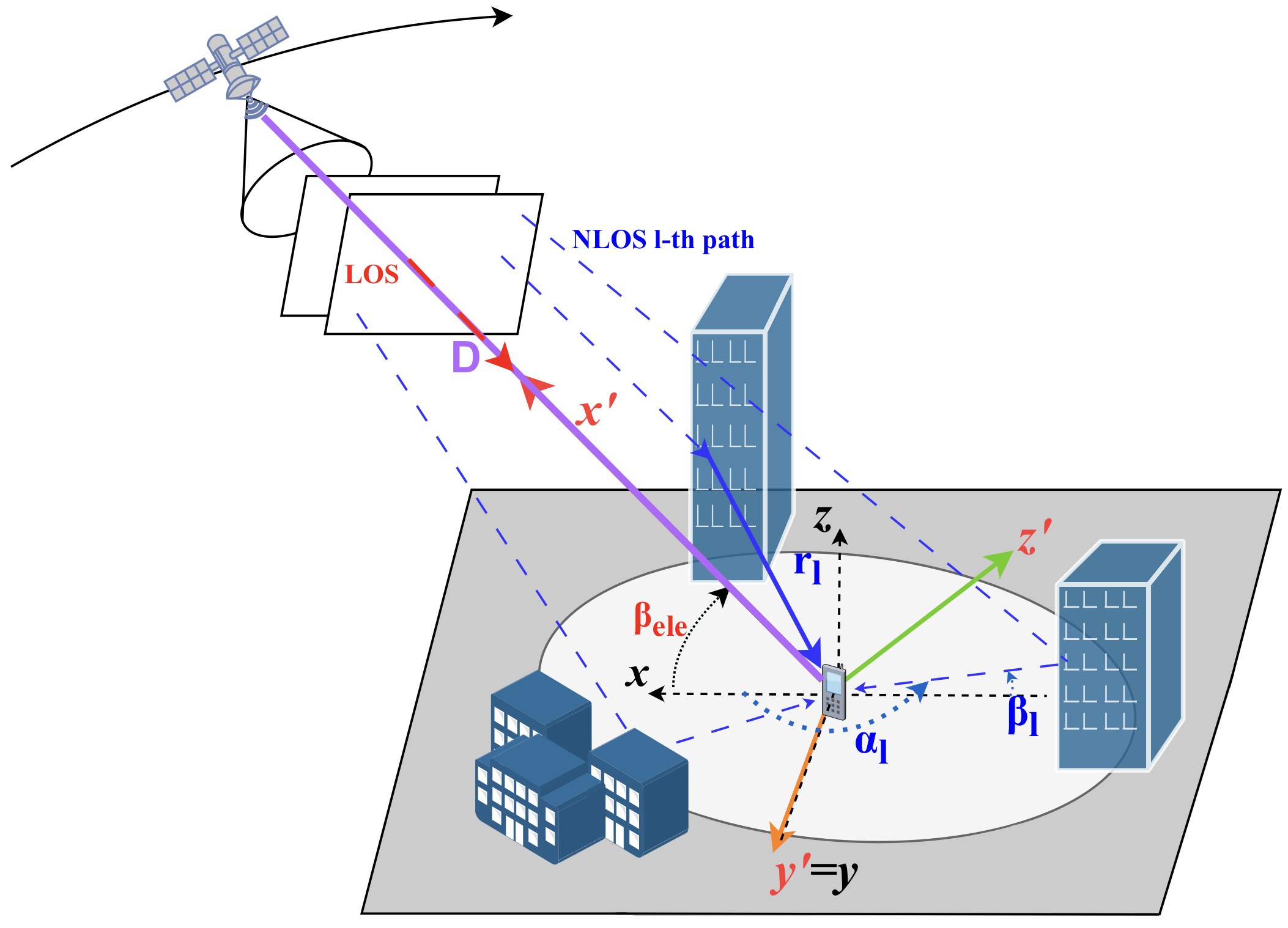}}
    \caption{The geometry of satellite-to-ground scatter model}
    \label{fig1}
\end{figure}

Consider a static receiver in an environment where a transmitted signal reflects off numerous nearby (to the terrestrial receiver) scatterers, each contributing a ray component as in Fig.~\ref{fig1}. 
 The \(l\)-th ray arrives at the receiver with an azimuth AoA $\alpha_l$, an elevation AoA \(\beta_l\), and a distance $r_l$ between scatter and user. The radial instantaneous relative velocity between the user and the satellite is denoted by \(v=\frac{dD}{dt}\), where $D$ is the slant distance. The contribution of \(l\)-th multipath ray with delay $\tau_l$ is expressed as:
\begin{equation}
    E_l(t)=A_le^{-j(2\pi f_l(t-\tau_l)+\Tilde{\phi}_l )}\, ,
\end{equation}
where $A_l=A_0g(\alpha_l,\beta_l)$ is the gain, with $A_0$ representing the amplitude of the incident plane wave and $g(\alpha_l,\beta_l)$ represents the Rx antenna beamforming. The phase of each ray is random $\Tilde{\phi_l}$. The frequency shift $f_l$ for the $l$-th ray due to instantaneous Doppler $f_d=\frac{v}{\lambda}$ may be written as 
\begin{equation}
f_l = f_d \cos\alpha_l \cos \beta_l
\label{eq:doppler}
\end{equation}
The total received E-field at time $t$ is given by
\begin{equation}
E(t)=\sum_{l=1}^{L}E_l(t)  
= \sum_{l=1}^{L}A_l \cdot e^{-j(2\pi f_l (t-\tau_l) + \Tilde{\phi}_l)},
\end{equation}
Suppose the smallest delay \(\tau_0=D/c_0\) is contributed by the line-of-sight (LoS) ray, where $c_0$ is the light speed. The relative delay \(\Delta \tau_l = \tau_l - \tau_0\) for $l$-th path is negligibly small compared to the symbol duration. Suppose channel is wide-sense stationary uncorrelated scattering (WSSUS) random process over time, with scattering components uncorrelated across distinct paths. Then the autocorrelation function of the net received E-field is:
\begin{equation}
\resizebox{0.48\textwidth}{!}{%
$\begin{aligned}
&R_E(\tau) = \mathbb{E}\{ E(t) \cdot E^*(t + \tau) \} \label{eq:RE1_def} \\
&= \sum_{l=1}^{L} \sum_{m=1}^{L} \mathbb{E} \left\{ A_l A_m^* e^{j [2\pi f_m(t+\tau-\tau_m) + 2\pi f_l( \tau_l - t) + (\tilde{\phi}_m - \tilde{\phi}_l)]} \right\} \\
&= \sum_{l=1}^{L} \sum_{m=1}^{L} \mathbb{E}\{ A_l A_m^*  e^{j2\pi [f_m(t+\tau-\tau_m) + f_l( \tau_l - t)]}\}  \mathbb{E}\{ e^{j (\tilde{\phi}_m - \tilde{\phi}_l)} \} 
\end{aligned}$
}
\end{equation}
Given WSSUS, $\mathbb{E}\{ e^{j (\Tilde{\phi}_m - \Tilde{\phi}_l)} \}=\delta (l-m) $. Now
\begin{equation}
\resizebox{0.48\textwidth}{!}{%
$\begin{aligned}
&R_E(\tau) = \sum_{l=1}^{L} |A_l|^2 e^{j 2\pi f_l \tau} = \sum_{l=1}^{L} |A_0|^2 |g(\alpha_l,\beta_l)|^2 e^{j 2\pi f_l \tau}  \\
&= \sum_{l=1}^{L} |A_0|^2 |G(\alpha_l,\beta_l)| e^{j 2\pi f_l \tau} \label{eq:re} \\
&\approx |A_0|^2 \int_{\alpha_l} \int_{\beta_l} G(\alpha_l,\beta_l) \, p(\alpha_l,\beta_l) \, e^{j 2\pi f_d \tau \cos\alpha_l \cos\beta_l} \, d\alpha_l \, d\beta_l. 
\end{aligned}$
}
\end{equation}
in the limit that the number of multipath components is large, where \(p(\alpha_l,\beta_l)\) represents the joint PDF of AoA. $G(\alpha_l,\beta_l) = | g(\alpha_l,\beta_l )|^2$ is the antenna power gain for given direction. With normalizing $|A_0|^2=1$, and isotropic antenna that $G(\alpha_l,\beta_l)=1$, $R_E(\tau)$ becomes:
\begin{equation}
    R_E(\tau)=\int_{\alpha_l} \int_{\beta_l} p(\alpha_l,\beta_l) \, e^{j2\pi f_d\tau\,\cos\alpha_l\,\cos\beta_l}\, d\alpha_l\, d\beta_l.
    \label{eq:re_tau}
\end{equation}

\subsection{Multipath Scattering Model Geometry}
\label{sec:model}
\begin{figure}[!thp]
    \centering
    \includegraphics[width=1\linewidth]{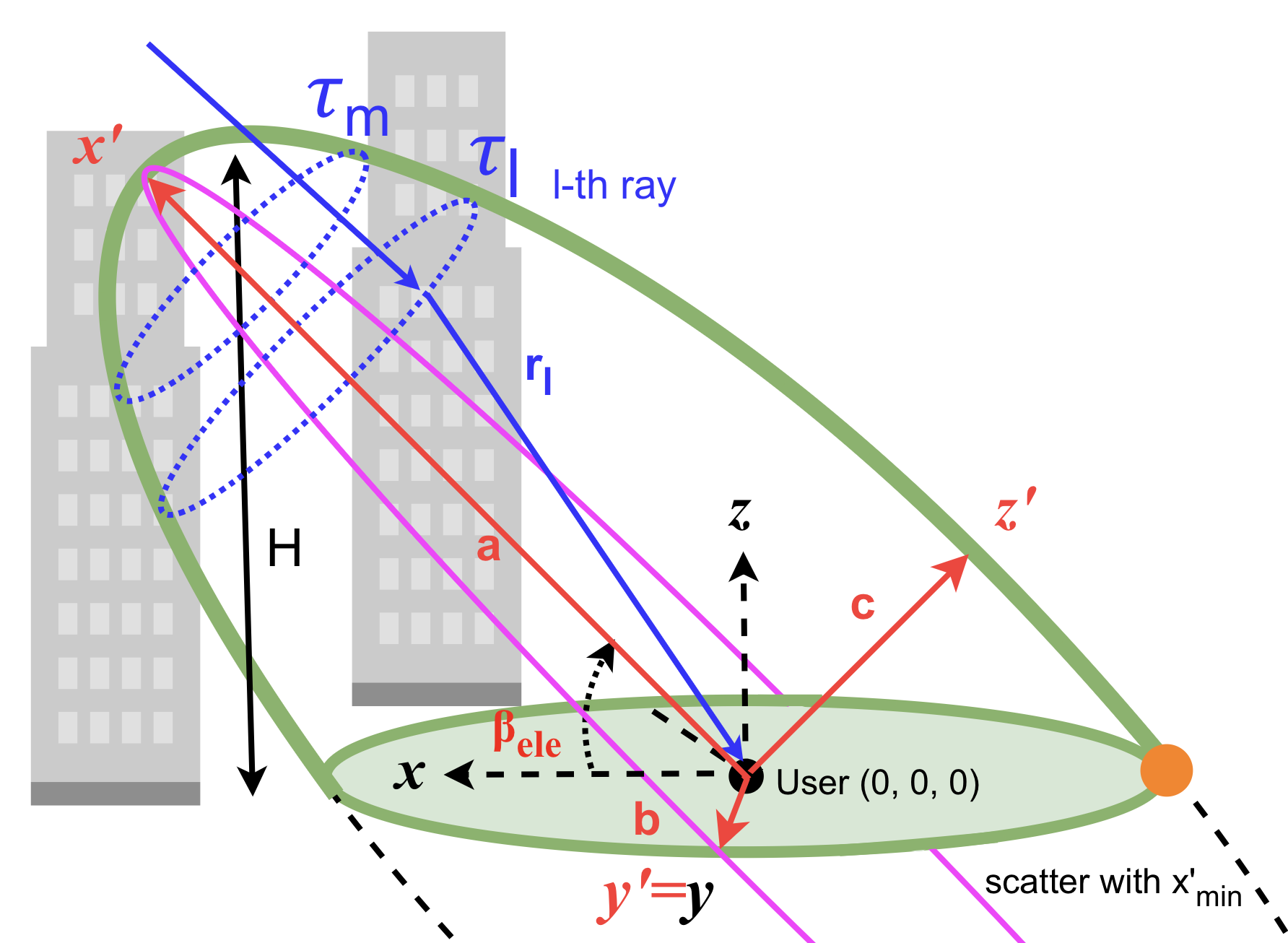}
    \caption{The proposed semi-ellipsoid scatter model}
    \label{fig2}
\end{figure}

To derive the joint PDF \( p(\alpha_l, \beta_l) \) for NLoS scenarios in \eqref{eq:re}, we introduce a volumetric spatial scatterer distribution that is uniform in the semi-ellipsoid volume shown in Fig.~\ref{fig2}. The semi-ellipsoid rotates about the \(y\)-axis by an angle \(\beta_{\text{ele}}\) such that its semi-major axis remains aligned with the LoS path to the satellite. The user locates at the origin \((0, 0, 0)\) of the global Cartesian coordinate system \((x, y, z)\), while the scatters are only in the upper semi-ellipsoid ($z>0$). The rotated coordinate system \((x', y', z')\), obtained by rotating the global frame, represents the ellipsoid as:
\begin{equation}
\frac{{x'}^2}{a^2} + \frac{{y'}^2}{b^2} + \frac{{z'}^2}{c^2} \leq 1,
\end{equation}
where \(a\), \(b\), and \(c\) denote the lengths of the semi-major and the two semi-minor axes, respectively.
The transformation from the global coordinates \((x, y, z)\) to the rotated coordinates \((x', y', z')\) is given by: 
\begin{equation}
\begin{pmatrix}
x' \\
y' \\
z'
\end{pmatrix}
=
\begin{pmatrix}
\cos\beta_{ele} & 0 & \sin\beta_{ele} \\
0 & 1 & 0 \\
-\sin\beta_{ele} & 0 & \cos\beta_{ele}
\end{pmatrix}
\begin{pmatrix}
x \\
y \\
z
\end{pmatrix}.
\label{matrix}
\end{equation}
With $r_l=\sqrt{x'^2+y'^2+z'^2}$, the conversion from Cartesian coordinate to spherical coordinate:
\begin{equation}
    \begin{pmatrix}
    x \\
    y \\
    z
    \end{pmatrix}=
    \begin{pmatrix}
    r_l\,cos\alpha_l\, cos\beta_l \\
    r_l\,sin\alpha_l\, cos\beta_l \\
    r_l\,sin\beta_l
    \end{pmatrix}
    \label{eq:spherical}
\end{equation}
 The propagation delay $\tau_l$ from satellite to user via a secondary scattering path is given by:
\begin{align}
\tau_l &= \resizebox{0.43\textwidth}{!}{\(\displaystyle \frac{\sqrt{(D-x')^2 + y'^2 + z'^2} + r_l}{c_0} = \frac{\sqrt{D^2 - 2Dx' + r_l^2} + r_l}{c_0}\)}
\end{align}
Since $D \, >> \, x', y', z, r_l$, $\sqrt{D^2 - 2Dx' + r_l^2}$ can be approximated using a Taylor expansion. Rewriting as $D\sqrt{1 + \left(-\frac{2x'}{D} + \frac{r_l^2}{D^2}\right)}$ and applying the first-order binomial approximation $\sqrt{1+\epsilon} \approx 1 + \frac{\epsilon}{2}$ with $\epsilon = -\frac{2x'}{D} + \frac{r_l^2}{D^2}$, we get:
\begin{equation}
\sqrt{D^2 - 2Dx' + r_l^2} \approx D \left[ 1 - \frac{x'}{D} + \frac{r_l^2}{2D^2} \right] \approx D - x'.
\end{equation}
implying $\tau_l=(D-x'+r_l)/c_0$, where $D-x'$ approximates the distance between the satellite and the scatterer. The {\em relative} delay $\Delta\tau_l$ for the $l$-th path equals:
\begin{equation}
\resizebox{0.48\textwidth}{!}{%
$\begin{aligned}
    \Delta\tau_l &= \frac{D - x' + r_l - D}{c_0} = \frac{r_l - x'}{c_0} \\
    &= \frac{r_l}{c_0} \left[1 - \left( \cos\alpha_l \cos\beta_l \cos\beta_{\mathrm{ele}} + \sin\beta_l \sin\beta_{\mathrm{ele}} \right)\right] \\
    &= \sqrt{x'^2 + y'^2 + z'^2} - x' = \sqrt{x'^2 + (\sqrt{y'^2 + z'^2})^2} - x' \label{eq:dtau}
\end{aligned}$
}
\end{equation}
Equation\eqref{eq:dtau} implies that any ($l$-th) ray reflected by the scatterers lying on a ring in the \(y'\)-\(z'\) plane experiences the same delay \(\tau_l\), where \(\tau_l\) is the function of the radius $\sqrt{y'^2+z'^2}$ of the ring. The delay \(\tau_m\) corresponds to another scatter ring located at a different distance \(D - x'\) from that of \(\tau_l\).
\subsection{Maximum Relative Delay and Maximum Building Height}
The maximum relative delay $\Delta\tau_{max}$ occurs at minimum $x'$ when $\alpha_l=-\pi$ and maximum $r_l$, i.e.
\begin{align}
x' &= x \cos \beta_{ele} + z \sin \beta_{ele} \\
   &= r_l \cos \alpha_l \cos \beta_l \cos \beta_{ele} + r_l \sin \beta_l \sin \beta_{ele} \\
   &= \resizebox{0.44\textwidth}{!}{\(\displaystyle \frac{r_l}{2} \left[ (\cos \alpha_l + 1) \cos(\beta_l - \beta_{ele}) + (\cos \alpha_l - 1) \cos(\beta_l + \beta_{ele}) \right]\)}
\end{align}
Then minimum $x'_{min}$ equals $-r_l cos(\beta_l-\beta_{ele})$. Further,  maximum $r_l$ can be obtained by expressing the rotated ellipsoid in spherical coordinates:
\begin{align}
&\Bigg[ \frac{\left( \cos\alpha_l \cos\beta_l \cos\beta_{ele} + \sin\beta_l \sin\beta_{ele} \right)^2}{a^2}  
+ \frac{(\sin\alpha_l \cos\beta_l)^2}{b^2} \notag \\
&+ \frac{\left( -\cos\alpha_l \cos\beta_l \sin\beta_{ele} + \sin\beta_l \cos\beta_{ele} \right)^2}{c^2)} \Bigg]r^2  = 1\,.
\end{align}
The maximum $r_l$ in a given direction  $r_{max}$ corresponds to a point on the surface of the ellipsoid:
\begin{align}
&r_{\text{max}}(\alpha_l, \beta_l) = \Bigg[ \frac{\left( \cos\alpha_l \cos\beta_l \cos\beta_{ele} + \sin\beta_l \sin\beta_{ele} \right)^2}{a^2} + \label{eq:rmax} \\  
&\resizebox{0.48\textwidth}{!}{$\frac{(\sin\alpha_l \cos\beta_l)^2}{b^2} + \frac{\left( \sin\beta_l \cos\beta_{ele}-\cos\alpha_l \cos\beta_l \sin\beta_{ele} \right)^2}{c^2} \Bigg]^{-\frac{1}{2}}$} \nonumber
\end{align}
% For $\alpha_l=-\pi$:
% \begin{equation}
%     r_{max}(-\pi,\beta_l)=\left( \frac{\cos^2(\beta_l - \beta_{ele})}{a^2} + \frac{\sin^2(\beta_l - \beta_{ele})}{c^2} \right)^{-1/2}
% \end{equation}
With \eqref{eq:dtau}, the minimum $x'$ occurs when $\alpha_l=-\pi, \beta_l=0$, represented by the orange dot in Fig.~\ref{fig2}:
\begin{equation}
    x'_{min} = -\cos\beta_{ele} \left( \frac{\cos^2\beta_{ele}}{a^2} + \frac{\sin^2\beta_{ele}}{c^2} \right)^{-1/2}
    \label{eq:x'}
\end{equation}
Substituting \eqref{eq:x'} into \eqref{eq:dtau}, maximum relative delay equals:
\begin{equation}
    \Delta\tau_{max} = \frac{1}{c_0} (1 + \cos\beta_{\text{ele}}) \left[ \frac{\cos^2\beta_{\text{ele}}}{a^2} + \frac{\sin^2\beta_{\text{ele}}}{c^2} \right]^{-1/2}
    \label{eq:tau_max}
\end{equation}
By considering the maximum building height $H$ and \eqref{eq:tau_max}, $a$ and $c$ can be solved. Using \eqref{eq:spherical}, the location of any point on the ellipsoid surface in the \(x'-z'\) plane is given by \((x', y', z') = (a \cos\beta_l, 0, c \sin\beta_l)\). The altitude of the point in the $x-y-z$ global Cartesian coordinate is:
\begin{align}
&z = a \cos\beta_l \sin\beta_{ele} + c \sin\beta_l \cos\beta_{ele} =\\
&= \resizebox{0.46\textwidth}{!}{$\sqrt{(a \sin\beta_{ele})^2 + (c \cos\beta_{ele})^2} \cos\left(\beta_l - \arctan\left(\frac{c \cos\beta_{ele}}{a \sin\beta_{ele}}\right)\right)$}
\nonumber
\end{align}

The corresponding maximum height is achieved when \(\beta_l = \arctan\left(\frac{c \cos\beta_{\text{ele}}}{a \sin\beta_{\text{ele}}}\right)\), given by:
\begin{equation}    
H = \max(|z|) = \sqrt{(a \sin\beta_{ele})^2 + (c \cos\beta_{ele})^2}.
\label{eq:h}
\end{equation}
% From \eqref{eq:h}, it follows that $c=H$ at \(\beta_{ele} = 0^\circ\) and \(a = H\)  at \(\beta_{ele} = 90^\circ\). 
Thus the semi-minor axis length $c$ can be written as 
\begin{align}
    c &= \sqrt{\frac{H^2 - a^2 \sin^2 \beta_{ele}}{\cos^2 \beta_{ele}}},  && 0^\circ \leq \beta_{ele} < 90^\circ 
    \label{eq:c1} 
\end{align}
Note that as $\beta_{ele} \rightarrow 90^\circ$, $ a \, \rightarrow H$ and thus $ c \, \rightarrow \, c_0\Delta\tau_{max}$. 
By giving the $H$ and $\Delta\tau_{max}$ the length of semi-major/semi-minor axes $a,c$ can be determined using \eqref{eq:c1} and \eqref{eq:tau_max}. Finally, the length of semi-minor axes $b$ can be obtained for a specified RMS delay spread.
\begin{align}
       &\sigma_\tau = \sqrt{\mathbb{E} \left[(\tau_l - \bar{\tau})^2\right]} = 
       \frac{1}{c_0}\sqrt{\mathbb{E}\left[(r_l-x')^2\right]-\mathbb{E}\left[r_l-x' \right]^2}.
       \label{eq:sigma_tau}
\end{align}

% According to \cite{8411465},\cite{8787874}, the maximum relative delay of aerial platform is usually within $0.01 \mu s$ to $5 \mu s$ depending on the environment and the carrier frequency, from observed measurements of the power delay profile. \\
\subsection{RMS Delay Spread}
 
 We next derive the RMS delay spread statistics for our model. The volume of scatterer semi-ellipsoid equals $\frac{2}{3}\pi abc$, hence the uniform PDF of scatterers over the half volume is
$f(x',y',z')=\frac{1}{V}=\frac{3}{2\pi abc }$. The Jacobian matrix \(J\) for coordinate conversion from $(x',y',z')$ to $(r_l,\alpha_l,\beta_l)$ equals $J = \frac{\partial(x',y',z')}{\partial(r_l,\alpha_l,\beta_l)}$, resulting in the absolute value of its determinant $\left| J \right| = r_l^2\,\cos\beta_l$, straightforwardly.  Using \eqref{eq:dtau}, the mean of $r_l-x'$ equals:
\begin{align}
&\mathbb{E}\left[r_l - x'\right] 
= \frac{1}{\frac{2}{3} \pi abc} \int_V (r_l - x') dV \\
&= \frac{1}{\frac{2}{3} \pi abc} \int_0^{2\pi} \int_0^{\frac{\pi}{2}} \int_0^{r_{\max}} r_l \left[1 - \left( \cos\alpha_l \cos\beta_l \cos\beta_{\mathrm{ele}} \right. \right. \nonumber\\
&\quad \left. \left. + \sin\beta_l \sin\beta_{\mathrm{ele}} \right)\right]  r_l^2 \cos\beta_l \, dr_l \, d\beta_l \, d\alpha_l  \\
&=\frac{3}{8 \pi abc} \int_0^{2\pi} \int_0^{\frac{\pi}{2}} \left[1 - \left( \cos\alpha_l \cos\beta_l \cos\beta_{\mathrm{ele}} \right. \right. \\
&\quad \left. \left. + \sin\beta_l \sin\beta_{\mathrm{ele}} \right)\right] \cos\beta_l \cdot r_{\max}^4(\alpha_l, \beta_l) \, d\beta_l \, d\alpha_l. \nonumber
\end{align}
The second moment of $r_l-x'$ is:
\begin{align}
&\mathbb{E}\left[(r_l - x')^2\right] 
= \frac{1}{\frac{2}{3} \pi abc} \int_V (r_l - x')^2 \, dV \\
&= \frac{1}{\frac{2}{3} \pi abc} \int_0^{2\pi} \int_0^{\frac{\pi}{2}} \int_0^{r_{\max}} 
\left[r_l - r_l \left( \cos\alpha_l \cos\beta_l \cos\beta_{\mathrm{ele}} \right. \right. \\
&\quad \left. \left. + \sin\beta_l \sin\beta_{\mathrm{ele}} \right)\right]^2 \cdot r_l^2 \cos\beta_l \, dr_l \, d\beta_l \, d\alpha_l \nonumber \\
&= \frac{3}{10 \pi abc} \int_0^{2\pi} \int_0^{\frac{\pi}{2}} 
\left[1 - \left( \cos\alpha_l \cos\beta_l \cos\beta_{\mathrm{ele}} \right. \right. \notag \\
&\quad \left. \left. +\ \sin\beta_l \sin\beta_{\mathrm{ele}} \right)\right]^2 \cdot \cos\beta_l \cdot r_{\max}^5(\alpha_l, \beta_l) \, d\beta_l \, d\alpha_l. 
\end{align}

\subsection{Implementation Flow of S2G channel model}
\label{sec:example}
\begin{figure}
    \centering
    \includegraphics[width=1\linewidth]{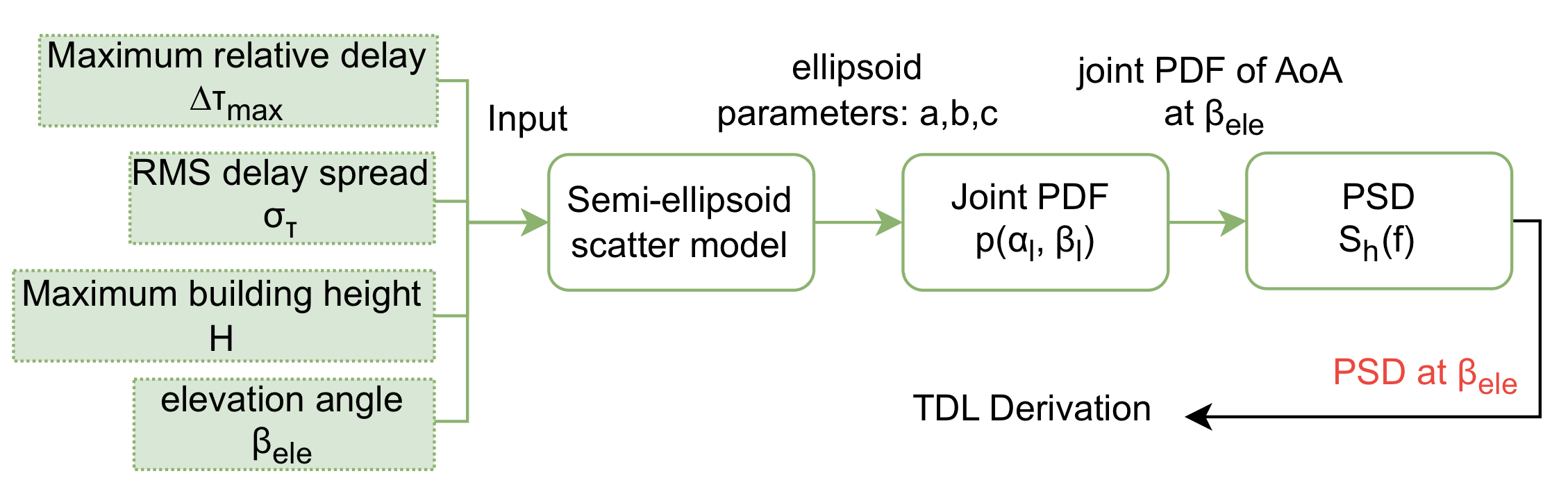}
    \caption{Flowchart of channel model implementation}
    \label{fig:flowchart}
\end{figure}

The ellipsoid scatter model at given $\beta_{ele}$ is driven by the RMS delay spread, H, max relative delay, which are defined in \eqref{eq:sigma_tau}, \ref{eq:c1}, \ref{eq:tau_max}. The flowchart in Fig.~\ref{fig:flowchart} illustrates the process of deriving a final Tapped Delay Line (TDL) channel model. In this paper, we derive the PSD of the received signal based on semi-ellipsoid scatter model characterized by joint PDF of scatterer distribution in AoA azimuth and elevation. 
The code used in generating the PSD results can be obtained @ Github repo below\footnote{\url{https://github.com/jessest94106/NTN_Channel_Model.git}}. 

\begin{figure}
    \centering
    \includegraphics[width=1\linewidth,height=0.48\textwidth]{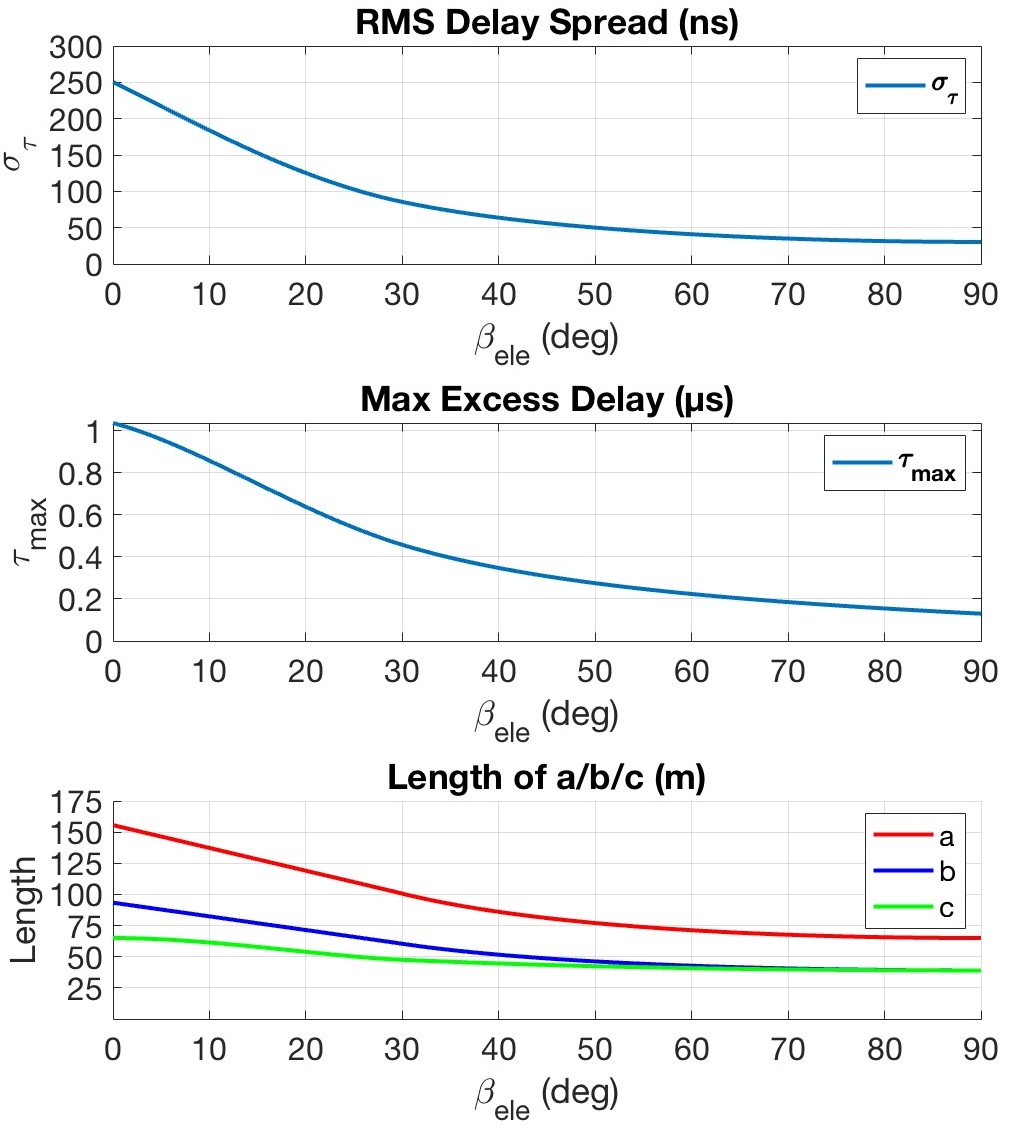}
    \caption{Length of a/b/c with given $H=65$ meters, RMS delay spread and max relative delay}
    \label{fig:ds}
\end{figure}

{\em Example}: The semi-major and semi-minor ellipsoid lengths $a,b,c$ can now be found, given $H$, RMS delay spread and max relative delay and is shown in Fig.\ref{fig:ds}. For convenience in subsequent computation, the max relative delay is set at the value that keeps $a$ and $b$ in constant ratio $b=a\sqrt{1-0.8^2}=0.6a$. 
Based on the ray-tracing model\cite{ray} and measurement results \cite{ka_mea}, the cumulative distribution function (CDF) of RMS delay spread for LEO-to-ground channel in NLOS urban environments is shown to range from 30 ns (min) to 250 ns (max). However, the trend for RMS delay spread as a function of $\beta_{ele}$ is not specified. Guidance for this can be developed via statistical analysis of the delay spread model in \cite{3gpp2020nr801}\cite{3GPP_R1-1807673}. The mean and standard deviation of delay spread decrease as $\beta_{ele}$ depending on the scattering environment and carrier frequency. This reduction is rapid at low elevation angles but slower at mid-to-high elevation angles. The realization of RMS delay spread in Table $\ref{tab:delay_spread}$ at given $\beta_{ele}$ is used for numerically computing $a,b,c$ at $1^\circ$ interval of $\beta_{ele}$ with linear interpolation.
\begin{table}[hb!]
\centering
\begin{tabular}{|c|c|}
\hline
\textbf{Elevation Angle (°)} & \textbf{RMS Delay Spread (ns)} \\
\hline
0   & 250 \\
10  & 183.7667 \\
20  & 125.1762 \\
30  & 85.4138 \\
40  & 63.7133 \\
50  & 50.0438 \\
60  & 40.9588 \\
70  & 34.9798 \\
80  & 31.5052 \\
90  & 30 \\
\hline
\end{tabular}
\caption{RMS Delay Spread vs. $\beta_{ele}$}
\label{tab:delay_spread}
\end{table}

\subsection{Azimuth and Elevation Angle - Joint \& Marginal PDF}
\label{sec:PDF}
With the $a,b,c$ obtained in \ref{sec:example}, the joint PDF can be computed.
The joint PDF in range, azimuth, and elevation planes is given by
\begin{align}
&f(r_l,\alpha_l,\beta_l)= r_l^2\cos\beta_l\, f(x',y',z')  
 \label{eq:jpdf_fixed} = \frac{3r_l^2\cos\beta_l}{2\pi abc}.
\end{align}
Integrating over \(r_l\) yields the joint pdf at given $\beta_{ele}$:
\begin{align}
&p(\alpha_l,\beta_l) 
= \int_{0}^{r_{max}} f(r_l,\alpha_l,\beta_l) \, dr_l = \frac{r_{max}^3(\alpha_l,\beta_l) \cos\beta_l}{2\pi abc}.
\label{eq:p()}
\end{align}

\begin{figure}[!thb]
    \centering
    \includegraphics[width=1\linewidth, height=0.37\textwidth]{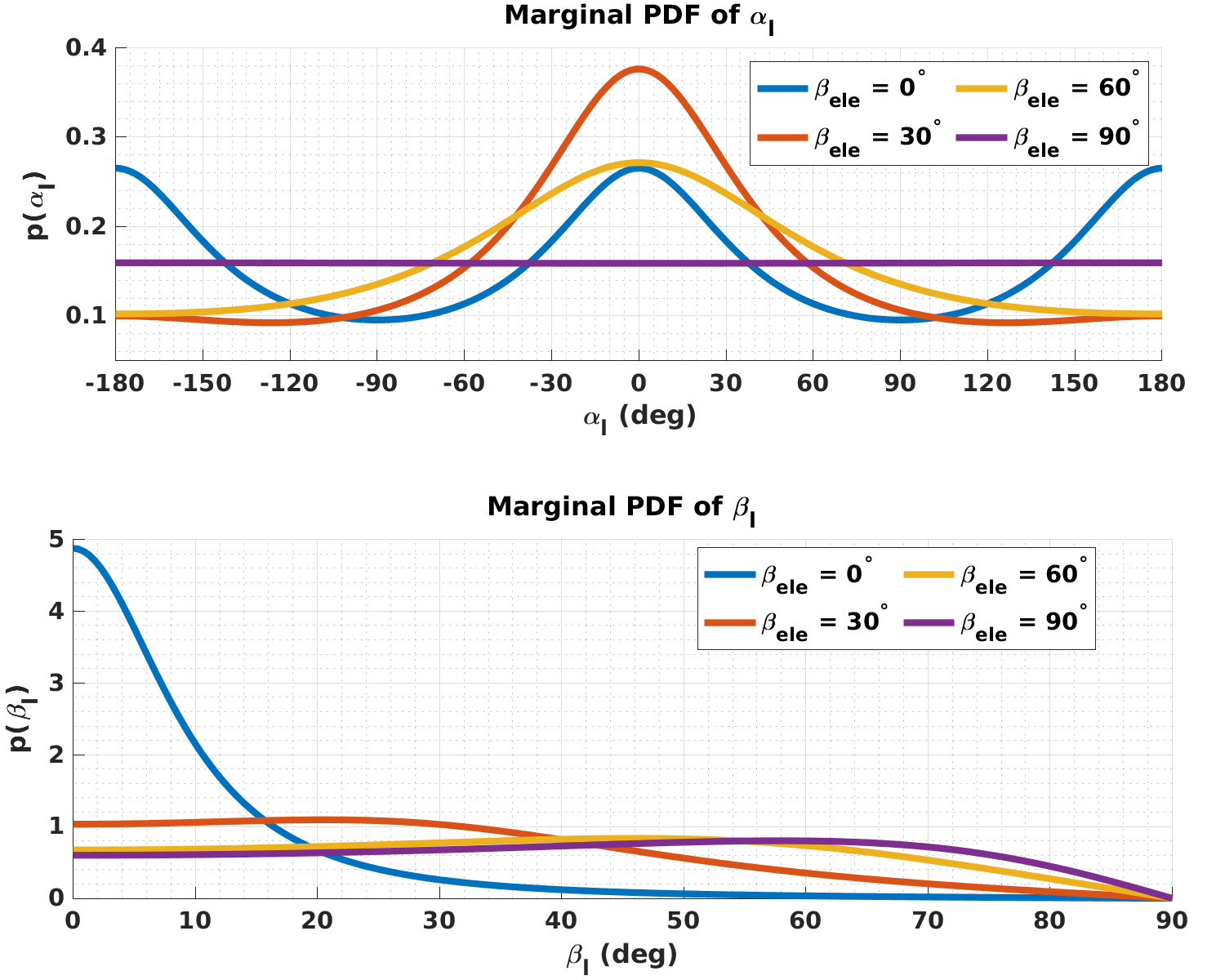}
    \caption{$p(\alpha_l)$ and $p(\beta_l)$ at a given elevation angle.}
    \label{fig:PDF_beta}
\end{figure}

The marginal PDFs of azimuth and elevation AoA are obtained by integrating \eqref{eq:p()} over the complementary variable.
In Fig.\ref{fig:PDF_beta}. With \(\Delta\tau_{max}\) as defined in Sec.~\ref{sec:example}, the AoA PDF peaks at \(\alpha_l = 0^\circ\) and \(\alpha_l = 180^\circ\) when the elevation angle \(\beta_{ele}=0^\circ\). The elevation AoA PDF peaks at $\beta_{ele}=0^\circ$ implies that most signal propagation is horizontal given the truth that $\beta_{ele}=0^\circ$. 
% exhibiting a trend similar to that observed in \cite{3GPP_R1-1807673} and \cite{doi:https://doi.org/10.1002/0470841524.ch5}. 
As the \( \beta_{ele} \) increases, the multipath contribution becomes more uniform horizontally, causing the azimuth AoA PDF to gradually flatten.
Meanwhile, the peak of the elevation AoA PDF shifts toward higher elevation angles, due to stronger vertical multipath contributions from tall buildings.

\subsection{General PSD}
To further discuss the relationship between PDF and PSD, the variables in \eqref{eq:re} are substituted with $u = \cos\alpha_l\in[-1, 1]$ and $v = \cos\beta_l\in[0,1]$.
With the corresponding derivatives $d\alpha_l = -\frac{du}{\sqrt{1-u^2}}$ and $ d\beta_l = -\frac{dv}{\sqrt{1-v^2}}$, the joint PDF from \eqref{eq:p()} in \( (u, v) \)-space becomes:
\begin{align}
&p_{u,v}(u, v) = p(\arccos u, \arccos v) \cdot \frac{1}{\sqrt{1 - u^2} \sqrt{1 - v^2}}
\end{align}
The PSD is given by the Fourier transform of $R_E(\tau)$ in \eqref{eq:re_tau}:
\begin{align}
    S(f) &= \int_{-\infty}^{\infty} R_E(\tau) e^{-j 2\pi f \tau} \, d\tau \nonumber\\
    &=\int_{-\infty}^{\infty} \left[ \int_{u} \int_{v} p_{u,v}(u, v) e^{j 2\pi f_d \tau u v} \, du \, dv \right] e^{-j 2\pi f \tau} \, d\tau \nonumber\\
    &=2\pi\int_{u} \int_{v} p_{u,v}(u, v)\delta\left(2\pi f_d uv - 2\pi f\right) \, du \, dv
    \label{eq:sf}
\end{align}
%The inner integral over \( v \):
%\[
%\int_{v} p_{u,v}(u, v) \delta(\omega_d u v - 2\pi f) \, dv
%\]
Let \( g(v) = 2\pi f_d u v - 2\pi f \), so \( g'(v) = 2\pi f_d u \), and \( v = \frac{f}{f_d u} \) when \( g(v) = 0 \). Using the delta function property:
\begin{equation}
\int_{v} p_{u,v}(u, v) \delta(g(v)) \, dv = \frac{p_{u,v}\left( u, \frac{f}{f_d u} \right)}{|g'(v)|} = \frac{p_{u,v}\left( u, \frac{ f}{f_d u} \right)}{|2\pi f_d u|}
\end{equation}
Thus:
\begin{equation}
    S(f) = \int_{u} \frac{p_{u,v}\left( u, \frac{f}{f_d u} \right)}{|f_d u|} \, du \, ,
    \label{eq:sf2}
\end{equation}
which shows that the final PSD only depends on the given joint PDF $p_{u,v}(u,v)$. It shows that the PSD can be understood intuitively through the given joint PDF.

\section{PSD of the Proposed Channel Model}
\label{sec:PSD}
Based on the PDF in Sec.\eqref{sec:PDF}, the PSD is numerically computed in MATLAB using \eqref{eq:sf}. The MATLAB \texttt{trapz} function is employed with 1000 grid points for numerical integration, while the PDF is evaluated using the \texttt{integral} function. For simplicity, both the PSD and frequency are presented on a normalized scale.
% Figure~\ref{fig:PSD2} illustrates the PSD with different $e_b$ at $\beta_{ele} = 60^\circ$ and positive $f_d$. When \( e_b = 0 \), the model simplifies to the same spherical scatter model as in Aulin's~\cite{aulin} and Janaswamy's~\cite{3dspheroid} models. The resulting PSD stays U-shaped and constant across \( \beta_{ele} \), misaligning with changing scatter geometry in the S2G channel.
% % \begin{figure}[!b]
% %     \centering
% %     \includegraphics[ width=\linewidth]{PSDe.png}
% %     \caption{PSD at different $e_b$ with \( \beta_{ele} = 60^\circ \)}
% %     \label{fig:PSD2}
% % \end{figure}
% For \( e_b > 0 \), the PSD skews toward positive frequencies, as a nonzero \( e_b \) stretches the illuminated scatterer area to the LoS direction, increasing multipath contributions in the positive frequency region of the PSD. Such behavior is consistent with ray-tracing trends reported in~\cite{3GPP_R1-1807673}. 

\begin{figure}[!b]
    \centering
    \includegraphics[width=1\linewidth,height=0.3\textwidth]{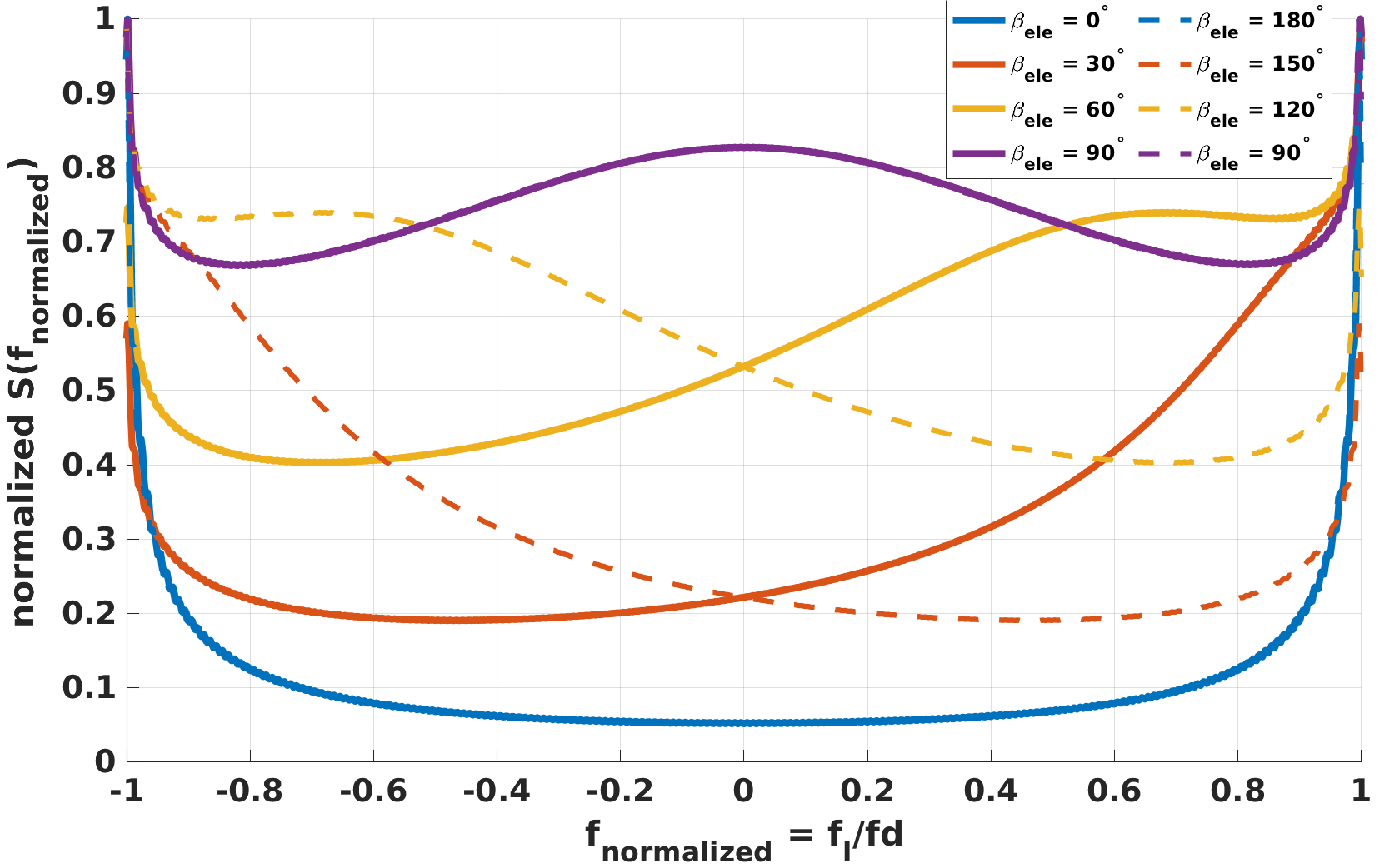}
    \caption{PSD at different elevation angles }
    \label{fig:PSD1}
\end{figure}

The elevation angle $\beta_{ele} \, \in \, [0^\circ, 90^\circ] $ yield $f_d > 0$, while $\beta_{ele} \, \in \, [90^\circ, 180^\circ]$ lead to $f_d < 0$ and are plotted as solid (dashed) lines in Figure~\ref{fig:PSD1}, respectively. When $\beta_{ele} = 0^\circ$, multipath components are centered at azimuth angles of $0^\circ$ and $180^\circ$ for the AoA. Given that the Doppler shift frequency of the incident wave is proportional to \(\cos\alpha_l \cdot \cos\beta_l\) as per \eqref{eq:doppler}, the resulting PSD exhibits higher values at the edges compared to Aulin's symmetric U-shaped PSD~\cite{aulin}. 
Since \(D \gg r_l\), the delay spread  \(\Delta\tau_l \, \approx \, (r_l - x')/c_0\). The farthest scatterer from the satellite (\(x'_{min}<0\)) corresponds to the maximum delay spread \(\Delta\tau_{max}\). The AoA of scatterers corresponding to maximum delay spread \(\Delta\tau_{max}\) is non-isotropic; more scatterers are located in the azimuth angle range \(\alpha_l \in [-90^\circ, 90^\circ]\), resulting in an azimuth AoA PDF concentrated within \([-90^\circ, 90^\circ]\). It leads to more multipath components with \(\cos\alpha_l > 0\) while \(\cos\beta_l \geq 0\), contributing to positive \(f_l = f_d \cos\alpha_l \cos\beta_l\). Thus the PSD skews toward positive frequencies as $\beta_{ele}<90^\circ$. If the scatters corresponding to maximum delay spread \(\Delta\tau_{max}\) are isotropic, which occurs when \(D \not\gg r_l\), i.e. the terrestrial scenario, it reduces to uniform spherical scatterer model assumptions as in Aulin's~\cite{aulin} and Janaswamy's~\cite{3dspheroid}. 
When $\beta_{ele} \rightarrow 90^\circ$, the vertical distribution of multipath components increasingly dominates the PSD. The azimuth AoA PDF becomes progressively uniform while the peaks of elevation AoA PDF shift toward higher $\beta_l$. Consequently, at $\beta_{ele} = 90^\circ$, the PSD exhibits peaks at both the center and the edges. Finally note that for $\beta_{ele} \in [90^\circ, 180^\circ]$, as the satellite moves away, the PSD is the symmetric flip of PSD with $\beta_{ele} \in [0^\circ, 90^\circ]$.
\begin{figure}[!tb]
    \centering
    \includegraphics[width=1\linewidth,height=0.3\textwidth]{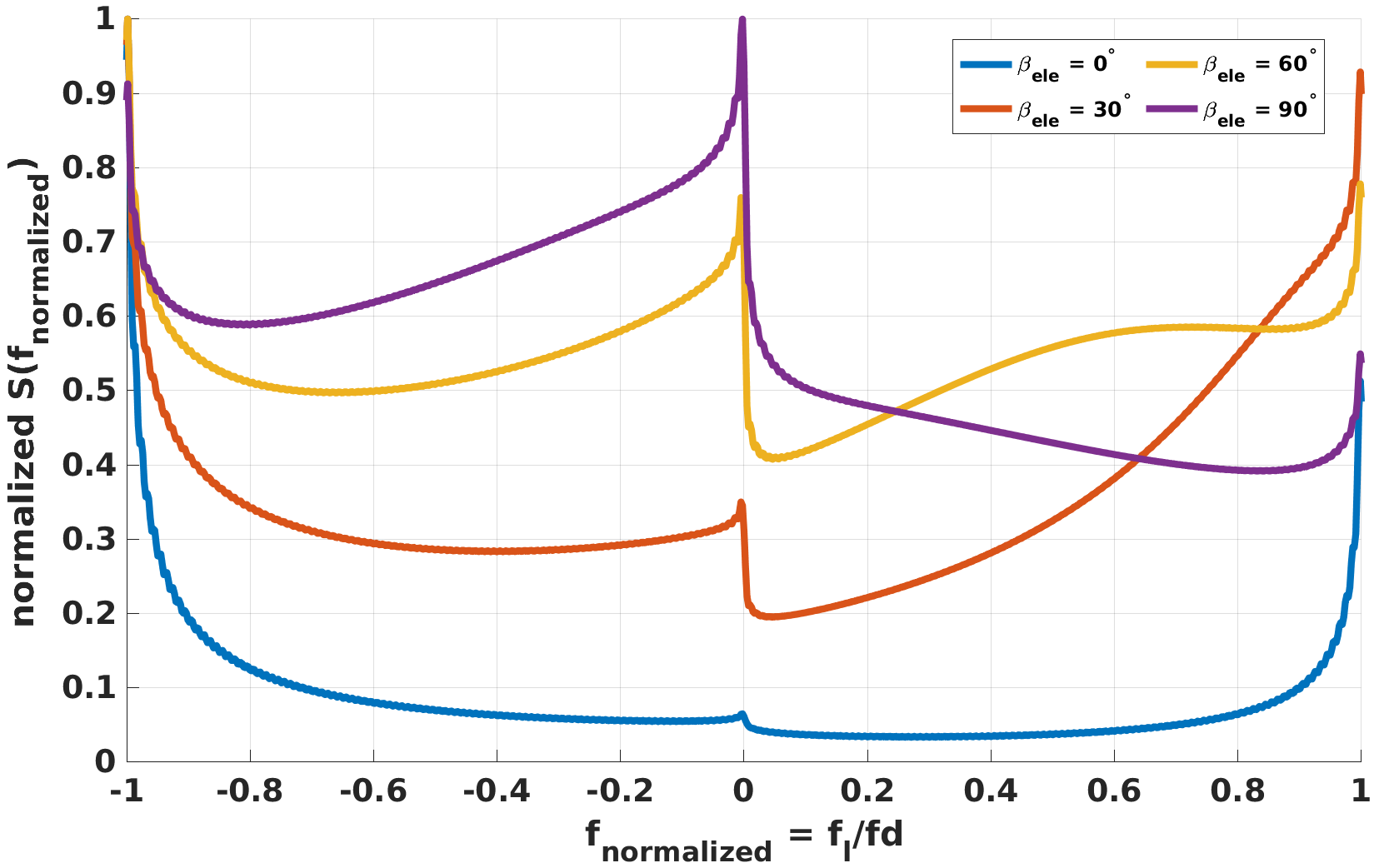}
    \caption{PSD at given elevation angle with $\alpha_l\in[0^\circ,270^\circ]$}
    \label{fig:PSD_sp}
\end{figure}

{\em Impact of Azimuth PDF Truncation}: Special cases of PSD occur when the azimuth PDF support is truncated, which may result from asymmetries in spatial distribution of scatterers (buildings) or in Rx antenna pattern. The discontinuous U-shaped PSD, first analyzed in \cite{53g} for a 2-D case with uniform and truncated azimuth AoA PDFs, has the discontinuity position governed by the azimuth PDF boundary. In~\cite{fingerprint}, a discontinuous U-shaped PSD was observed in the LEO downlink, but its dependence on \( \beta_{ele} \) remains unclear. While \cite{53g} provides a 2D example with \( \alpha_l \) ranging from \( 0^\circ \) to \( 270^\circ \), a 3D PSD example with varying \( \beta_{ele} \) is shown in Fig.~\ref{fig:PSD_sp}. The discontinuity at $f_l/f_d = 0$ with the PSD for $ f = 0^+$ being lower than $f=0^-$ occurs because scatters within $\alpha_l\in[270^\circ,90^\circ]$ contribute to $f_l>0$ while scatters within $\alpha_l\in[90^\circ, 270^\circ]$ contribute to $f_l<0$. By limiting \( \alpha_l \) to \( [0^\circ, 270^\circ] \) excludes scatterer contributions from the \( [270^\circ, 360^\circ] \) quadrant, removing some of the multipath components with positive $f_l$.
% As the elevation and azimuth AoA PDFs are non-uniform, truncating their range distorts the PSD shape.
At low \(\beta_{ele}\), the azimuth AoA PDF is concentrated at \(\alpha_l = 0^\circ\), causing more Doppler frequencies to be centered at \(f_l/f_d = \pm 1\). Consequently, when \(\beta_{ele} = 0^\circ\), most multipath components are centered at \(f_l/f_d = \pm 1\), resulting in a minimal PSD difference across the discontinuity at \(f_l/f_d = 0\). As \(\beta_{ele}\) increases, the azimuth AoA PDF flattens, shifting more \(|f_l|\) toward 0, which increases the PSD discontinuity at \(f_l/f_d = 0\).

\section{Conclusion}
This paper introduces a novel geometry-inspired LEO S2G downlink channel model for NLoS environments, focusing on how scattering and satellite geometry shapes resulting Doppler and the PSD. Our semi-ellipsoid scatter model derives joint AoA PDFs as functions of \( \beta_{ele} \) and captures how PSD varies with elevation angle. The discontinuous azimuth PDF model reveals its impact on the PSD across varying \( \beta_{ele} \). The PSD will be further developed into a time-domain channel model using the flowchart presented in Fig. \ref{fig:flowchart}, as future work.

% If you are using \IEEEpubid in a two-column layout, remember to call \IEEEpubidadjcol in the second column.
% \IEEEpubidadjcol

\bibliographystyle{IEEEtran} 
\bibliography{reference}           

\end{document}